\documentclass{aa}
\usepackage{graphicx}
\usepackage{epsfig}
\usepackage{natbib}
\usepackage{rotating}
\usepackage{txfonts}
\bibpunct{(}{)}{;}{a}{}{,}
\begin{document}

\title{The ELODIE survey for northern extra-solar planets\\
III. Three planetary candidates detected with ELODIE\thanks{
Based on observations made with the {\footnotesize ELODIE} echelle spectrograph mounted on the 1.93--m Telescope 
at the Observatoire de Haute-Provence ({\footnotesize CNRS}) and with the {\footnotesize CORALIE} echelle spectrograph mounted 
on the 1.2--m Euler Swiss Telescope at {\footnotesize ESO}--La\,Silla Observatory.} \thanks{The 
{\footnotesize ELODIE} and {\footnotesize CORALIE} measurements discussed in this paper are only available in 
electronic form at the {\footnotesize CDS} via anonymous ftp to {\tt 
cdsarc.u-strasbg.fr (130.79.128.5)} or via 
{\tt http://cdsweb.u-strasbg.fr/cgi-bin/qcat?J/A+A/}}}

\author{D.~Naef\inst{1} \and M.~Mayor\inst{1} \and J.L.~Beuzit\inst{2} \and
C.~Perrier\inst{2} \and D.~Queloz\inst{1} \and J.P.~Sivan\inst{3} \and S.~Udry\inst{1}
}

\institute{Observatoire de Gen\`eve, 51 ch. des Maillettes, 
CH--1290 Sauverny, Switzerland \and
Laboratoire d'Astrophysique, Observatoire de Grenoble, 
Universit\'e J. Fourier, BP 53, F--38041 Grenoble, France \and
Observatoire de Haute-Provence, F--04870 St-Michel L'Observatoire, 
France 
}

\offprints{Dominique Naef,
\email{Dominique.Naef@obs.unige.ch}}

\date{Received / Accepted}
\abstract{We present our {\footnotesize ELODIE} radial-velocity measurements of 
\object{{\footnotesize HD}\,74156} and \object{14\,Her} (\object{{\footnotesize HD}\,145675}). These stars 
exhibit low-amplitude radial-velocity variations induced by the presence of low-mass companions. 
The radial-velocity data of \object{{\footnotesize HD}\,74156} reveal the presence of two planetary companions: 
a 1.86\,M$_{\rm Jup}$ planet on a 51.64--d  orbit and a 
6.2\,M$_{\rm Jup}$ planet on a long-period ($\simeq$\,5.5 yr) orbit. Both orbits are fairly eccentric 
($e$\,=\,0.64 and 0.58).
The 4.7\,M$_{\rm Jup}$ companion to \object{{\footnotesize HD}\,145675} has a long period (4.9\,yr) and a moderately 
eccentric orbit ($e$\,=\,0.34). We detect an additional linear radial-velocity trend superimposed to the periodic signal 
for this star. We also compute updated orbital solutions for \object{{\footnotesize HD}\,209458} and \object{51\,Peg} 
(\object{{\footnotesize HD}\,217014}). Finally, we present our {\footnotesize ELODIE} radial-velocity data and orbital 
solutions for 5 stars known to host planetary companions: 
\object{Ups\,And} (\object{{\footnotesize HD}\,9826}), \object{55\,Cnc} (\object{{\footnotesize HD}\,75732}), \object{47\,UMa} (\object{{\footnotesize HD}\,95128}), 
\object{70\,Vir} (\object{{\footnotesize HD}\,117176}) and \object{{\footnotesize HD}\,187123}. We confirm the previously published orbital 
solutions for \object{Ups\,And}, \object{70\,Vir} and \object{{\footnotesize HD}\,187123}. Our data are not sufficient for fully 
confirming the orbital solutions for \object{55\,Cnc} and \object{47\,UMa}.
}

\titlerunning{The ELODIE survey for northern extra-solar planets III}
\authorrunning{D.~Naef et al.}
\maketitle

\section{Introduction}\label{intro}
The {\sl ELODIE survey for northern extra-solar planets}, a programme aiming at detecting and characterising  
planetary companions in orbit around solar-type stars of the solar neighbourhood, started in 1994. This survey uses 
the {\footnotesize ELODIE} echelle spectrograph \citep{Baranne96} mounted on the Cassegrain focus of the 1.93--m Telescope 
of the Observatoire de Haute-Provence ({\footnotesize CNRS}, France). 

The original \citet{Mayorcssss} sample consisted of 142 stars, including \object{51\,Peg}, the first solar-type star known to host 
a planetary companion \citep{Mayor51peg}. 
A new sample of 330 stars was defined in 1996. The new sample, the observing procedure and the achieved precision are 
described in details in \citet[][ Paper\,I]{PerrierELODIE1}. The complete list of planets detected with 
{\footnotesize ELODIE} can also be found in Paper\,I.

In this paper, we present two systems hosting long-period planets.
We present in Sect.~\ref{newplanets} our {\footnotesize ELODIE} radial-velocity measurements for the solar-type stars 
\object{{\footnotesize HD}\,74156} and \object{14\,Her}. 
Updated velocities and orbits for \object{{\footnotesize HD}\,209458} and \object{51\,Peg} are presented 
in Sect.~\ref{updatedplanets}. Section~\ref{confirmedplanets} contains the description of the 
{\footnotesize ELODIE} radial-velocity data and orbital solutions for \object{Ups\,And}, \object{55\,Cnc}, \object{47\,Uma}, \object{70\,Vir} and 
\object{{\footnotesize HD}\,187123}. Our results are summarized in Sect.~\ref{conc}.

\section{New ELODIE planet candidates}\label{newcands}

\subsection{Stellar characteristics}\label{stars}
\begin{table}[th!]
\caption{
\label{tabstar}
Observed and inferred stellar parameters for \object{HD\,74156} and \object{14\,Her}. 
}
\begin{tabular}{ll|r@{  $\,\pm\,$  }lr@{  $\,\pm\,$  }l}
\hline\hline
\noalign{\vspace{0.025cm}}
\multicolumn{2}{c}{}                                     & \multicolumn{2}{c}{\bf \object{{\footnotesize HD}\,74156}} & \multicolumn{2}{c}{\bf 14\,Her}\\
\noalign{\vspace{0.025cm}}
\hline
\noalign{\vspace{0.025cm}}
HIP                         &                            & \multicolumn{2}{c}{42723}                         & \multicolumn{2}{c}{79248}\\
$Sp.\,Type$                 &                            & \multicolumn{2}{c}{G0$^\dagger$}                   & \multicolumn{2}{c}{K0V}          \\
$m_{\rm V}$                 &                            & \multicolumn{2}{c}{7.61}                          & \multicolumn{2}{c}{6.61}         \\
$B-V$                       &                            & 0.585     & 0.014                                 & 0.877     & 0.006                \\
$\pi$                       & (mas)                      & 15.49     & 1.10                                  & 55.11     & 0.59                 \\
$Distance$                  & (pc)                       & 64.56     & $^{4.93}_{4.28}$                      & 18.15     & $^{0.20}_{0.19}$     \\
$\mu _{\alpha}\cos(\delta)$ & (mas yr$^{-1}$)            & 24.96     & 1.20                                  & 132.52    & 0.52                 \\
$\mu _{\delta}$             & (mas yr$^{-1}$)            & $-$200.48 & 0.81                                  & $-$298.38 & 0.55                 \\
$M_{\rm V}$                 &                            & \multicolumn{2}{c}{3.56}                          & \multicolumn{2}{c}{5.32}         \\
$B.C.$                      &                            & \multicolumn{2}{c}{$-$0.031}                      & \multicolumn{2}{c}{$-$0.208}     \\
$L$                         & (L$_{\sun}$)               & \multicolumn{2}{c}{3.06}                          & \multicolumn{2}{c}{0.71}         \\
$T_{\rm eff}$               & ($\degr$K)                 & 6105     & 50                                     & 5255     & 50                    \\
$\log g$                    & (cgs)                      & 4.40     & 0.15                                   & 4.40     & 0.15                  \\
$\xi _{\rm t}$              & (km\,s$^{-1}$)             & 1.36     & 0.10                                   & 0.68     & 0.10                  \\
$[$Fe/H$]$                  &                            & 0.15     & 0.06                                   & 0.51     & 0.06                  \\ 
$M_{\ast}$                  & (M$_{\sun}$)               & \multicolumn{2}{c}{1.27}                          & \multicolumn{2}{c}{0.9}          \\
$v\sin i$                   & (km\,s$^{-1}$)             & 4.06     & 0.62                                   & \multicolumn{2}{c}{$<$1}         \\
$\log R^{'}_{\rm HK}$       &                            & \multicolumn{2}{c}{--}                            & \multicolumn{2}{c}{$-$5.07}      \\
$P_{\rm rot}$               & (d)                        & \multicolumn{2}{c}{--}                            & \multicolumn{2}{c}{41}           \\
$age _{\rm HK}$             & (Gyr)                      & \multicolumn{2}{c}{--}                            & \multicolumn{2}{c}{3.9}          \\
\noalign{\vspace{0.025cm}}
\hline
\noalign{\vspace{0.025cm}}
\end{tabular}

$^\dagger$ the rather large luminosity of this star probably reveals a slightly evolved status (IV-V luminosity class).
\\
\end{table}

In this section, we briefly describe the main properties of the stars hosting the new planet candidates detected with 
{\footnotesize ELODIE}, namely \object{{\footnotesize HD}\,74156} and \object{14\,Her}. The properties of the 
other stars appearing in this paper are described elsewhere.

The main characteristics of \object{{\footnotesize HD}\,74156} and  \object{14\,Her} are listed in 
Table~\ref{tabstar}.
Spectral Type, $m_{\rm V}$, $B-V$, $\pi$, $\mu _{\alpha}\cos(\delta)$ and $\mu _{\delta}$ are from the 
{\footnotesize HIPPARCOS} Catalogue \citep{ESA97}. The effective temperature $T_{\rm eff}$, the surface gravity $\log g$, 
the metallicity $[$Fe/H$]$, the microturbulance velocity $\xi_{\rm t}$ and the stellar masses $M_{\ast}$ 
are taken from \citet{Santosstat}. 
The bolometric correction $B.C.$ is 
computed from the effective temperature with the calibration in \citet{Flower96}.

The projected rotational velocity $v\sin i$ is derived from the {\footnotesize ELODIE} cross-correlation function widths using 
the calibration in \citet{Queloz98}. We only have a $v\sin i$ upper limit for \object{14\,Her} because this 
star does not exhibit any rotational broadening measurable with {\footnotesize ELODIE}. 
The $\log R^{'}_{\rm HK}$ chromospheric activity indicator for 
\object{14\,Her} is from \citet{Butler14Her}. No quantitative activity value is available for \object{{\footnotesize HD}\,74156}. 
We do not see any trace of emission in the core of the $\lambda$\,3968.5\,\AA\hspace{1mm}\ion{Ca}{ii} H line on 
our {\footnotesize ELODIE} coadded spectra. This star is thus probably not very active although moderately fast 
rotating. $P_{\rm rot}$ and $age({\rm HK})$ are derived from the activity indicator using the calibrations in 
\citet{Noyes84} and \citet{Donahue93}\footnote{also quoted in \citet{Henry96}}, respectively.

\object{{\footnotesize HD}\,74156} and \object{14\,Her} are not photometrically variable. 
Their measured {\footnotesize HIPPARCOS} scatter is 11 and 9 mmag, respectively. Moreover, \object{14\,Her} 
is flagged as constant star in this catalogue.

We have 3 adaptive optics high-angular resolution images of \object{14\,Her}. They were obtained 
between May 1997 and August 2001 using the {\footnotesize PUE'O} adaptive optics system 
mounted on the 3.6--m Canada-France Hawaii Telescope. 
No visual companion is detected around this star.

\subsection{Radial-velocity analysis}\label{newplanets}

\subsubsection{\object{HD\,74156} (\object{HIP\,42723})}\label{74156}

\begin{figure}
  \psfig{file=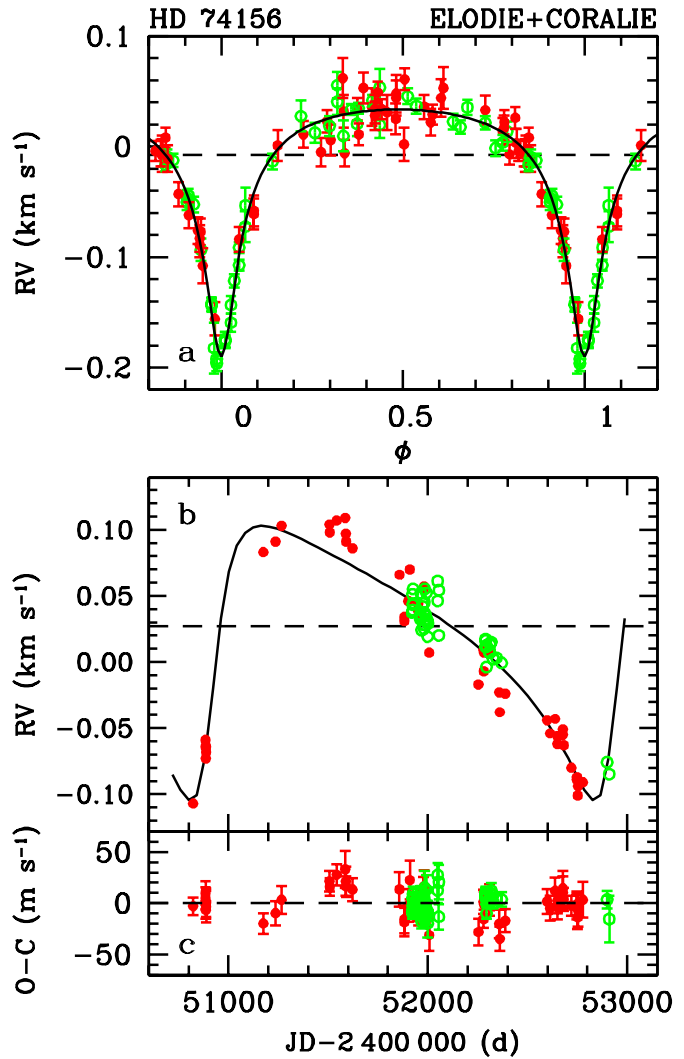,width=\hsize}\\
  \caption{
  \label{fig74156} ELODIE (filled dots) and CORALIE (open dots) radial-velocity data for \object{HD\,74156}.
  {\bf a:} Phase-folded velocities for the short-period planet.
  {\bf b:} Temporal velocities for the long-period planet.
  The residuals to the fitted 2--planet orbital solution are displayed in panel {\bf c}.
  }  
\end{figure}

\begin{table*}[t!]
\caption{
\label{tab74156}
Orbital solution obtained for \object{HD\,74156} by combining the ELODIE (E) and CORALIE (C) radial-velocity data.
$\Delta RV_{E-C}$ is the radial-velocity offset between the ELODIE and CORALIE
measurements. 
$\sigma_{\rm O-C}$ is the weighted rms of the residuals.
$\chi^{\rm 2}_{\rm red}$ is the reduced $\chi^{\rm 2}$ value ($\chi^{\rm 2}/\nu$ where $\nu$ is the number of degrees of freedom).
}
\begin{tabular}{ll|r@{ \,$\pm$\, }lr@{ \,$\pm$\, }l|r@{ \,$\pm$\, }l}
\hline\hline
\noalign{\vspace{0.05cm}}
                                    &                                  & \multicolumn{2}{c}{{\bf \object{HD\,74156\,b}}} & \multicolumn{2}{c|}{{\bf \object{HD\,74156\,c}}} & \multicolumn{2}{c}{{\bf \object{14\,Her\,b}}}\\[0.05cm]
$P$                                 & (d)                              & 51.643      & 0.011                    & 2025        & 11                        & 1796.4    & 8.3\\
$T$                                 & (JD$^{\dagger}$)                 & 51\,981.321 & 0.091                    & 50\,901     & 10                        & 49\,582.0 & 12.4\\
$e$                                 &                                  & 0.636       & 0.009                    & 0.583       & 0.039                     & 0.338     & 0.011\\
$w$                                 & ($\degr$)                        & 181.5       & 1.4                      & 242.4       & 4.0                       & 22.58     & 1.98\\ 
$K_{\rm 1}$                         & (m\,s$^{\rm -1}$)                & 112.0       & 1.9                      & 104.0       & 5.5                       & 90.3      & 1.0\\
Slope                               & (m\,s$^{\rm -1}$\,yr$^{\rm -1}$) & \multicolumn{4}{c|}{--}                                                          & 3.6       & 0.3\\
$a_{\rm 1} \sin i$                  & (10$^{\rm -4}$AU)                & 4.10        & 0.06                     & 157         & 6                         & 140       & 2\\
$f_{\rm 1}(m)$                      & (10$^{\rm -9}$M$_{\sun}$)        & 3.45        & 0.15                     & 126         & 14                        & 114       & 4\\
$m\sin i$                           & (M$_{\rm Jup}$)                  & 1.86        & 0.03                     & 6.17        & 0.23                      & 4.74      & 0.06\\
$a$                                 & (AU)                             & \multicolumn{2}{c}{0.294}              &\multicolumn{2}{c|}{3.40}                & \multicolumn{2}{c}{2.80}\\[0.05cm]
\hline
\noalign{\vspace{0.05cm}}
$\gamma$                            & (km\,s$^{\rm -1}$)               & \multicolumn{4}{c|}{3.840\,$\pm$\,0.003}                                         & $-$13.8226$^{\ddagger}$ & 0.0007\\
$\Delta RV_{EC}$                    & (m\,s$^{\rm -1}$)                & \multicolumn{4}{c|}{24.8\,$\pm$\,2.8}                                            & \multicolumn{2}{c}{--}\\
\hline
\noalign{\vspace{0.05cm}}
$N$                                 &                                  & \multicolumn{4}{c|}{95}                                                          & \multicolumn{2}{c}{119}\\ 
$N_{E}$                             &                                  & \multicolumn{4}{c|}{51}                                                          & \multicolumn{2}{c}{119}\\
$N_{C}$                             &				       & \multicolumn{4}{c|}{44}                                                          & \multicolumn{2}{c}{--}\\
$\sigma_{\rm O-C}$                  & (m\,s$^{\rm -1}$)                & \multicolumn{4}{c|}{10.6}                                                        & \multicolumn{2}{c}{11.3}\\
$\sigma_{\rm O-C, E}$               & (m\,s$^{\rm -1}$)                & \multicolumn{4}{c|}{14.0}                                                        & \multicolumn{2}{c}{11.3}\\
$\sigma_{\rm O-C, C}$               & (m\,s$^{\rm -1}$)                & \multicolumn{4}{c|}{8.9}                                                         & \multicolumn{2}{c}{--}\\
$\chi^{\rm 2}_{\rm red}$            &                                  & \multicolumn{4}{c|}{1.63}                                                        & \multicolumn{2}{c}{2.59}\\[0.05cm]
\hline
\end{tabular}

$^{\dagger}$=\,{\footnotesize JD}$-$2\,400\,000, $^{\ddagger}$ at {\footnotesize JD}\,=\,2\,451\,500
\end{table*}

The detection of a two-planet system orbiting \object{{\footnotesize HD}\,74156} 
was already announced by our team in April 2001 
(April 4$^{th}$ 2001 {\footnotesize ESO} Press Release).
The long-period of the outer required however further observations of this system.

We now have 51 {\footnotesize ELODIE} and 44 {\footnotesize CORALIE} radial-velocity measurements of 
\object{{\footnotesize HD}\,74156}. These velocities have been obtained between {\footnotesize JD}\,=\,2\,450\,823 (January 1998) and 
{\footnotesize JD}\,=\,2\,452\,910 (September 2003). The mean velocity uncertainty of these measurements is 12.7\,m\,s$^{\rm -1}$ 
for the {\footnotesize ELODIE} data and 8.5\,m\,s$^{\rm -1}$ for the {\footnotesize CORALIE} data.
The total measurement time span is 2087\,d.

The fitted orbital elements for {\footnotesize HD}\,74156
 are presented in Table~\ref{tab74156}.
Figure~\ref{fig74156} shows the fitted orbits and 
the residuals around the solution.  
Assuming a primary mass of 1.27\,M$_{\sun}$ \citep{Santosstat}, we compute the planetary minimum masses 
$m_{\rm b}\sin i$\,=\,1.86\,M$_{\rm Jup}$ 
and $m_{\rm c}\sin i$\,=\,6.17\,M$_{\rm Jup}$ as well as the semi-major axes 
$a_{\rm b}$\,=\,0.294\,AU and $a_{\rm c}$\,=\,3.40\,AU.

\subsubsection{\object{14\,Her} (\object{HD\,145675}, \object{HIP\,79248}, \object{GJ\,614})}\label{14her}

\begin{figure}
  \psfig{file=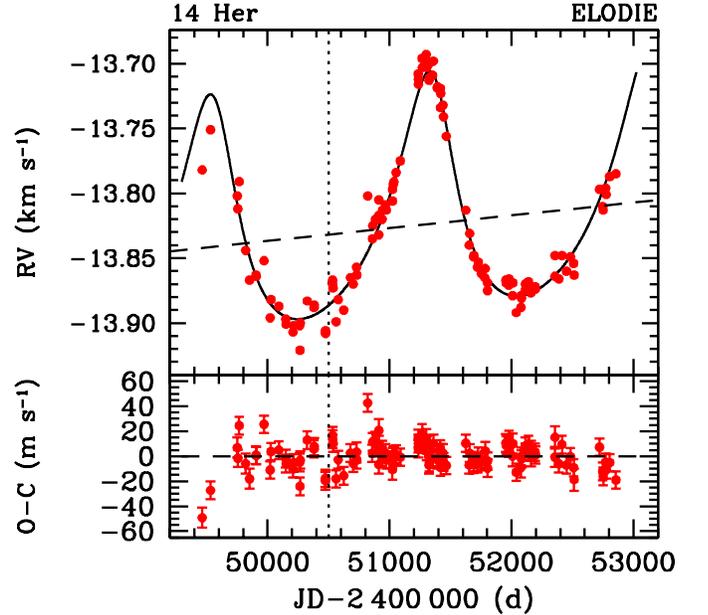,width=\hsize}\\
  \caption{
  \label{fig14her} ELODIE radial-velocity data and orbital solution for \object{14\,Her}. 
  {\bf Top:} Temporal velocities. The dashed line shows the drifting systemic velocity.
  {\bf Bottom:} Residuals to the fitted  orbit. We note the significantly larger scatter of the measurements taken before 
  the set up of the double scrambling device (indicated by the dotted line).
  }  
\end{figure}

The detection with {\footnotesize ELODIE} of a long-period planetary companion to \object{14\,Her} was announced during the 
{\sl Protostars \& Planets IV} conference by M.~Mayor  \citetext{1998, oral contribution, also in \citealt*{MarcyPPIV}}.
An updated orbital solution was presented by \citet{Udryman}. This detection was recently confirmed by \citet{Butler14Her}.

The orbital solution of Table~\ref{tab74156} was obtained by fitting a Keplerian orbit to the 119 {\footnotesize ELODIE} 
radial-velocity measurements. These measurements were obtained between {\footnotesize JD}\,=2\,449\,464 (April 1994) and 
{\footnotesize JD}\,=2\,452\,857 (August 2003). The time span of these data is though 3392\,d. Our velocities have a mean 
velocity uncertainty of 7.2\,m\,s$^{\rm -1}$.

The residuals to a simple Keplerian fit are abnormally large: 14\,m\,s$^{\rm -1}$. We thus computed another 
solution including an additional linear velocity drift. The fitted drift parameter is highly significant: 
3.6\,$\pm$\,0.3\,m\,s$^{\rm -1}$\,yr$^{\rm -1}$. The residuals around the Keplerian+drift solution are much 
smaller: $\sigma_{\rm O-C}$\,=\,11.3\,m\,s$^{\rm -1}$. This value is still a bit large compared to our uncertainties.

\begin{figure}[th!]
  \psfig{file=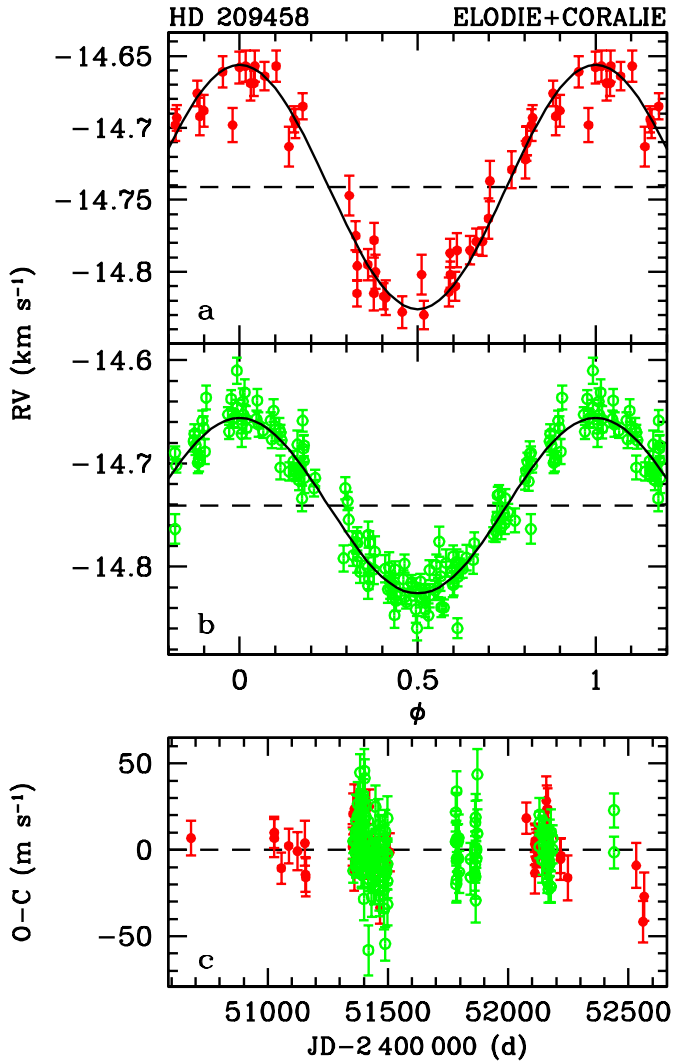,width=\hsize}\\
  \caption{
  \label{fig209458} Radial-velocity data and orbital solutions for \object{HD\,209458}. The measurements taken at or near the 
  transits have been removed.
  {\bf a:} ELODIE Phased-folded velocities and combined solution.
  {\bf b:} CORALIE Phased-folded velocities and combined solution.
  {\bf c:} ELODIE (filled dots) and CORALIE (open dots) residuals to the combined orbit. 
  }  
\end{figure}

Figure~\ref{fig14her} shows the {\footnotesize ELODIE} radial-velocity data, the fitted Keplerian + linear drift solution and 
the residuals to the solution. The residuals of the oldest measurements have clearly a higher scatter value. We computed 
an orbital solution using only the most recent observations (velocities taken after {\footnotesize JD}\,=2\,450\,500, the 
date of the set up of the double scrambling device). The 
solution we obtained in this case was compatible with the solution quoted in Table~\ref{tab74156} (slightly longer period 
$P$\,=1853\,$\pm$\,35\,d, same eccentricity and semi-amplitude) but with a significantly smaller residuals value: 
$\sigma_{\rm O-C}$\,=\,8.6\,m\,s$^{\rm -1}$ ($\chi^{\rm2}_{\rm red}$\,=\,1.54). The linear drift slope of this solution was 
slightly less significant: 3.9\,$\pm$\,1.1\,m\,s$^{\rm -1}$\,yr$^{\rm -1}$ which is not a surprise since only one 
radial-velocity minimum is present in the most recent data. 
The temporal coverage of these latter data approximately corresponds to the coverage of the \citet{Butler14Her} measurements. 
We also computed an orbital solution to this velocity set without including the drift parameter. The fitted parameters 
obtained in this case are in good agreement with the \citet{Butler14Her} values but the residuals are slightly larger 
than for the drift-included solution. This shows that the non detection of the velocity drift by these authors is 
probably due to their much shorter observing span (1882\,d). Moreover, their data only cover one velocity maximum and one 
minimum.

The computed minimum mass of \object{14\,Her\,b}, assuming $M_{\rm 1}$\,=\,0.9\,M$_{\sun}$ \citep{Santosstat}, is 4.74\,M$_{\rm Jup}$ 
 and its semi-major axis is 2.80\,AU.

\section{Updated orbits of known extra-solar planets}\label{updatedplanets}

In this section, we present updated orbital solutions for \object{{\footnotesize HD}\,209458} and \object{51\,Peg}. These two stars 
host planetary companions that were detected with {\footnotesize ELODIE} \citep{Mazeh2000, Mayor51peg}. 
The new orbital solutions are based on larger and more precise radial-velocity sets.

\subsection{\object{HD\,209458} (\object{HIP\,108859})}\label{hd209458}

\begin{table}[t!]
\caption{
\label{tab209458}
ELODIE (E) + CORALIE (C) updated orbital solution for \object{HD\,209458} and ELODIE updated orbital solution for \object{51\,Peg}
The parameters have the same definitions as in Table~\ref{tab74156}.
}
\begin{tabular}{ll|r@{\,$\pm$\,}lr@{\,$\pm$\,}l}
\hline\hline
\noalign{\vspace{0.05cm}}
\multicolumn{2}{c}{}                                                   & \multicolumn{2}{c}{{\bf \object{HD\,209458\,b}}} & \multicolumn{2}{c}{{\bf \object{51\,Peg\,b}}}\\
\hline
\noalign{\vspace{0.05cm}}								        
$P$                                 & (d)                              & 3.5246      & 0.0001           & 4.23077     & 0.00004\\
$T$                                 & (JD$^{\dagger}$)                 & 617.758     & 0.021            & 497.0       & 0.022\\
$e$                                 &                                  & \multicolumn{2}{c}{0 (fixed)}  & \multicolumn{2}{c}{0 (fixed)}\\
$w$                                 & ($\degr$)                        & \multicolumn{2}{c}{0 (fixed)}  & \multicolumn{2}{c}{0 (fixed)}\\ 
$K_{\rm 1}$                         & (m\,s$^{\rm -1}$)                & 85.1        & 1.0              & 57.3        & 0.8\\
$a_{\rm 1} \sin i$                  & (10$^{\rm -5}$AU)                & 2.75        & 0.03             & 2.23        & 0.03\\
$f_{\rm 1}(m)$                      & (10$^{\rm -10}$M$_{\sun}$)       & 2.25        & 0.07             & 0.82        & 0.03\\
$m\sin i$                           & (M$_{\rm Jup}$)                  & 0.699       & 0.007            & 0.468       & 0.007\\
$a$                                 & (AU)                             & \multicolumn{2}{c}{0.048}      & \multicolumn{2}{c}{0.052}\\[0.05cm]
\hline
\noalign{\vspace{0.05cm}}
$\gamma$                            & (km\,s$^{\rm -1}$)               & $-$14.741   & 0.002            & $-$33.2516  & 0.0006\\
$\Delta RV_{\rm EC}$                & (m\,s$^{\rm -1}$)                & 14.3        & 1.8              & \multicolumn{2}{c}{--}\\
\hline
\noalign{\vspace{0.05cm}}
$N$                                 &                                  & \multicolumn{2}{c}{187}        & \multicolumn{2}{c}{153}\\ 
$N_{\rm E}$                         &                                  & \multicolumn{2}{c}{46}         & \multicolumn{2}{c}{153}\\ 
$N_{\rm C}$                         &                                  & \multicolumn{2}{c}{141}        & \multicolumn{2}{c}{--}\\ 
$\sigma_{\rm O-C}$                  & (m\,s$^{\rm -1}$)                & \multicolumn{2}{c}{14.9}       & \multicolumn{2}{c}{11.8}\\
$\sigma_{\rm O-C, E}$               & (m\,s$^{\rm -1}$)                & \multicolumn{2}{c}{13.7}       & \multicolumn{2}{c}{11.8}\\
$\sigma_{\rm O-C, C}$               & (m\,s$^{\rm -1}$)                & \multicolumn{2}{c}{15.2}       & \multicolumn{2}{c}{--}\\
$\chi^{\rm 2}_{\rm red}$            &                                  & \multicolumn{2}{c}{2.36}       & \multicolumn{2}{c}{2.72}\\[0.05cm]
\hline
\noalign{\vspace{0.05cm}}
\end{tabular}

$^{\dagger}$=\,{\footnotesize JD}$-$2\,452\,000
\end{table}
\begin{figure}[th!]
  \psfig{file=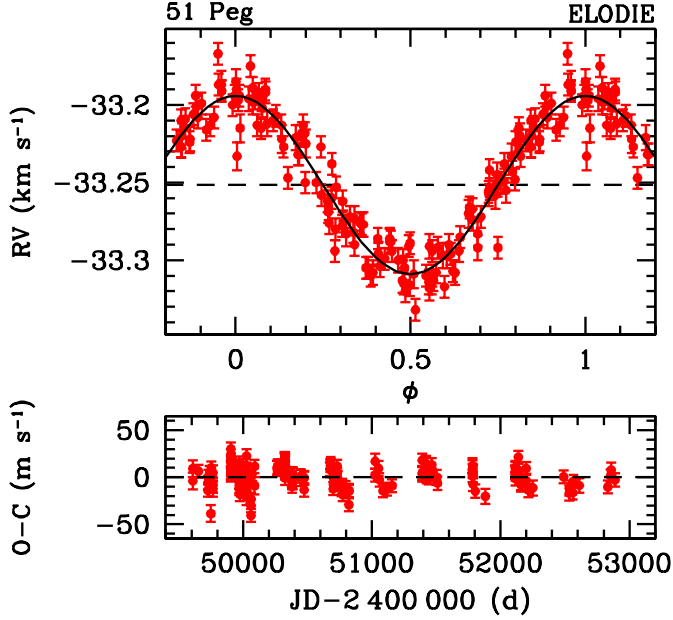,width=\hsize}\\
  \caption{
  \label{fig51peg} ELODIE radial-velocity data spanning over 3000 days and fitted orbital solutions for \object{51\,Peg}. 
  {\bf Top:} Phased-folded velocities.
  {\bf Bottom:} Residuals to the fitted orbit. 
  }  
\end{figure}

\begin{table}[th!]
\caption{
\label{tabupsand}
ELODIE orbital solutions for \object{Ups\,And} and \object{55\,Cnc}.
The parameters have the same definitions as in Table~\ref{tab74156}.
}
\begin{tabular}{ll|r@{\,$\pm$\,}l|r@{\,$\pm$\,}l}
\hline\hline
\noalign{\vspace{0.05cm}}
                                    &                                  & \multicolumn{2}{c|}{{\bf \object{Ups\,And\,b}}} &\multicolumn{2}{c}{{\bf \object{55\,Cnc\,b}}}\\[0.05cm]
$P$                                 & (d)                              & 4.61712     & 0.00009                  & 14.647     & 0.001\\
$T$                                 & (JD$^{\dagger}$)                 & 4.28        & 0.49                     & 0.80       & 0.48\\
$e$                                 &                                  & 0.020       & 0.023                    & 0.030      & 0.023\\
$w$                                 & ($\degr$)                        & 241.5       & 37.9                     & 63         & 12\\ 
$K_{\rm 1}$                         & (m\,s$^{\rm -1}$)                & 77.2        & 1.3                      & 78.3       & 1.8\\
$a_{\rm 1} \sin i$                  & (10$^{\rm -5}$AU)                & 3.28        & 0.05                     & 10.54      & 0.24\\
$f_{\rm 1}(m)$                      & (10$^{\rm -10}$M$_{\sun}$)       & 2.20        & 0.11                     & 7.28       & 0.50\\
$m\sin i$                           & (M$_{\rm Jup}$)                  & 0.75        & 0.01                     & 0.91       & 0.02\\
$a$                                 & (AU)                             & \multicolumn{2}{c|}{0.059}             & \multicolumn{2}{c}{0.115}\\[0.05cm]
                                    &                                  & \multicolumn{2}{c|}{{\bf \object{Ups\,And\,c}}} & \multicolumn{2}{c}{{\bf \object{55\,Cnc\,d}}}\\[0.05cm]
$P$                                 & (d)                              & 238.10      & 0.46                     & 4545       & 1421\\
$T$                                 & (JD$^{\dagger}$)                 & 159.4       & 8.0                      & 568        & 200\\
$e$                                 &                                  & 0.185       & 0.028                    & 0.24       & 0.13\\
$w$                                 & ($\degr$)                        & 214.2       & 11.4                     & 347        & 23\\ 
$K_{\rm 1}$                         & (m\,s$^{\rm -1}$)                & 62.9        & 1.7                      & 37.8       & 3.9\\
$a_{\rm 1} \sin i$                  & (10$^{\rm -2}$AU)                & 0.135       & 0.003                    & 1.54       & 0.56\\
$f_{\rm 1}(m)$                      & (10$^{\rm -9}$M$_{\sun}$)        & 5.85        & 0.43                     & 23         & 11\\
$m\sin i$                           & (M$_{\rm Jup}$)                  & 2.25        & 0.06                     & 2.89       & 0.47\\
$a$                                 & (AU)                             & \multicolumn{2}{c|}{0.821}             & \multicolumn{2}{c}{5.28}\\[0.05cm]
                                    &                                  & \multicolumn{2}{c|}{{\bf \object{Ups\,And\,d}}} & \multicolumn{2}{c}{}\\[0.05cm]
$P$                                 & (d)                              & 1319        & 18                       & \multicolumn{2}{c}{}\\
$T$                                 & (JD$^{\dagger}$)                 & $-$37       & 53                       & \multicolumn{2}{c}{}\\
$e$                                 &                                  & 0.269       & 0.036                    & \multicolumn{2}{c}{}\\
$w$                                 & ($\degr$)                        & 247.6       & 10.6                     & \multicolumn{2}{c}{}\\ 
$K_{\rm 1}$                         & (m\,s$^{\rm -1}$)                & 63.8        & 2.3                      & \multicolumn{2}{c}{}\\
$a_{\rm 1} \sin i$                  & (10$^{\rm -3}$AU)                & 7.45        & 0.25                     & \multicolumn{2}{c}{}\\
$f_{\rm 1}(m)$                      & (10$^{\rm -8}$M$_{\sun}$)        & 3.17        & 0.30                     & \multicolumn{2}{c}{}\\
$m\sin i$                           & (M$_{\rm Jup}$)                  & 3.95        & 0.13                     & \multicolumn{2}{c}{}\\
$a$                                 & (AU)                             & \multicolumn{2}{c|}{2.57}              & \multicolumn{2}{c}{}\\[0.05cm]
\hline
\noalign{\vspace{0.05cm}}
$\gamma$                            & (km\,s$^{\rm -1}$)               & $-$28.655    & 0.002                   & 27.252   & 0.009\\
\hline
\noalign{\vspace{0.05cm}}
$N$                                 &                                  & \multicolumn{2}{c|}{71}                & \multicolumn{2}{c}{48}\\ 
$\sigma_{\rm O-C}$                  & (m\,s$^{\rm -1}$)                & \multicolumn{2}{c|}{14.9}              & \multicolumn{2}{c}{9.0}\\
$\chi^{\rm 2}_{\rm red}$            &                                  & \multicolumn{2}{c|}{4.34}              & \multicolumn{2}{c}{1.99}\\[0.05cm]
\hline
\end{tabular}

$^{\dagger}$=\,{\footnotesize JD}$-$2\,450\,000
\end{table}

The detection of \object{{\footnotesize HD}\,209458\,b} was simultaneously announced by \citet{Mazeh2000} and \citet{Henry209458}. 
Since the discovery paper, we have obtained more radial-velocity measurements of \object{{\footnotesize HD}\,209458}. 
Moreover, the numerical template we use for computing the cross-correlation function has 
been changed in order to reduce the effect of telluric atmospheric lines \citep{Pepe2002}. Another effect of this new 
template is the change of the radial-velocity zero point.

We now have 46 {\footnotesize ELODIE} and 141 {\footnotesize CORALIE} radial-velocity measurements of 
\object{{\footnotesize HD}\,209458} spanning from {\footnotesize JD}\,=\,2\,450\,681 (August 1997) and 
{\footnotesize JD}\,=\,2\,452\,566 (October 2002). The mean velocity uncertainty of these measurements is 
11.0\,m\,s$^{\rm -1}$ with {\footnotesize ELODIE} and 10.2\,m\,s$^{\rm -1}$ with {\footnotesize CORALIE}.

\object{{\footnotesize HD}\,209458}\,b is known to be a transiting extra-solar planet \citep{Charbonneau2000}. It has also been 
shown by \citet{Queloztrans} that the passage of the planet in front of the star was also affecting the 
measured radial-velocity (departure from the Keplerian model). We thus have removed all the measurement taken 
at or near the phase of the transit (measurements with 
$\phi_{\rm trans}\,-\,$0.025\,$\leq$\,$\phi$\,$\leq$\,$\phi_{\rm trans}$\,+\,0.025). 

We first tried to fit an orbital solution with the eccentricity as a free parameter. The obtained value was very low: 
$e$\,=\,0.015\,$\pm$\,0.019. We then compared the residuals around this solution with the value obtained from a circular 
solution. We found the same residuals for both solutions. According to the \citet{Lucy} test, the probability of 
significant eccentricity is zero for these data. Monte-Carlo simulations performed with the {\footnotesize ORBIT} 
software \citep{Forveille1999} give the following 3\,$\sigma$ uncertainty for this parameter: +0.063/$-$0.067. An eccentricity larger 
than 0.08 can thus be rejected with a very high confidence level.

We present in Table~\ref{tab209458} the circular orbital solution fitted to the {\footnotesize ELODIE}+{\footnotesize CORALIE} radial 
velocities.

The velocities and orbital solution are displayed in Fig.~\ref{fig209458}. This Figure also shows the 
residuals to the fitted orbit. We note a spectacular improvement of the solution quality since the initial publication 
\citep{Mazeh2000}. It results from new template used for the cross-correlation \citep{Pepe2002}.

\citet{Wittenmyer} have recently presented updated high-precision period and transit time for \object{{\footnotesize HD}\,209458}. 
Their determinations, based on space and ground-based photometric measurements of transits, are:
$P$\,=\,3.5247542\,$\pm$\,0.0000004\,d and $T_{\rm trans}$\,=\,2\,452\,618.66774\,$\pm$\,0.000007. Our period is compatible  
with their value (1.5\,$\sigma$ away). Our transit time prediction obtained from the fitted time of maximum radial 
velocity ($T_{\rm trans}$\,=\,2\,452\,618.639\,$\pm$\,0.021) is 1.3\,$\sigma$ away from their value. 
We have computed an orbital solution with $P$ fixed to the \citet{Wittenmyer} value and $T$ fixed to the time of 
maximum velocity computed from their $T_{\rm trans}$. The obtained values for the three remaining free parameters are less 
than 1\,$\sigma$ away from the values in Table~\ref{tab209458}.

With a mass of 1.15\,M$_{\sun}$ for \object{{\footnotesize HD}\,209458} \citep{Santosstat}, we compute the planetary 
minimum mass and semi-major axis: $m_{\rm 2}\sin i$\,=\,0.699\,M$_{\rm Jup}$ and $a$\,=\,0.048\,AU.

\begin{figure}[th!]
  \psfig{file=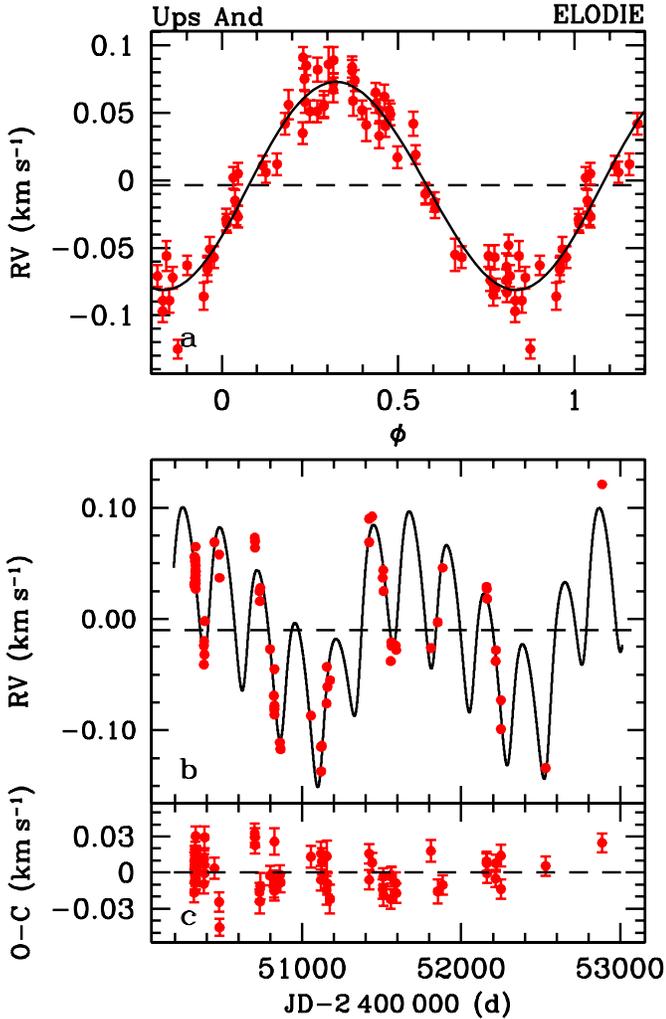,width=\hsize}\\
  \caption{
  \label{figupsand} ELODIE Radial-velocity data and orbital solution for \object{Ups\,And} 
  {\bf a:} Phased-folded velocities for \object{Ups\,And\,b}. The velocity 
  effect induced by the outer planets has been removed. 
  {\bf b:} Temporal velocities for \object{Ups\,And\,c} and d. The signal induced by the inner planet has been removed. 
  {\bf c:} Residuals to the fitted 3-planet orbital solution. The residuals value is larger than the expected value. 
  }  
\end{figure}

\subsection{\object{51\,Peg} (\object{HD\,217014}, \object{HIP\,113357}, \object{GJ\,882})}\label{51peg}

The detection of the first extra-solar planet around a Solar-type star, \object{51\,Peg\,b}, was published in \citet{Mayor51peg}.
This detection was rapidly confirmed by \citet{Marcy51peg}. Since the discovery, we have gathered many 
{\footnotesize ELODIE} radial-velocity measurements of \object{51\,Peg}.

The number of measurements for this star now amounts to 153. 
The time span of these observations is 3278\,days 
(from {\footnotesize JD}\,=\,2\,449\,610 (September 1994) to {\footnotesize JD}\,=\,2\,452\,888 (September 2003)) and their 
mean velocity uncertainty is 7.3\,m\,s$^{\rm -1}$.

The updated orbital solution for \object{51\,Peg} is presented in Table~\ref{tab209458}. 
As for {\footnotesize HD}\,209458, the \citet{Lucy} test shows that a fitted non-zero eccentricity is not 
significant. We thus fitted a circular orbit. Figure~\ref{fig51peg} shows the 
phase-folded velocities and fitted orbit together with the residuals to the solution.
The {\sl rms} of the residuals for this 
solution is 11.8\,m\,s$^{\rm -1}$, a value slightly smaller than the 13\,m\,s$^{\rm -1}$ obtained by \citet{Mayor51peg}.

Assuming a mass of 1.04\,M$_{\sun}$ for \object{51\,Peg} \citep{Santosstat}, we compute the planetary minimum mass and semi-major axis:
$m_{\rm 2}\sin i$\,=\,0.468\,M$_{\rm Jup}$ and $a$\,=\,0.052\,AU.

\section{Confirmations of planetary companions}\label{confirmedplanets}

In the same way as the presence of the companion to \object{51\,Peg} was confirmed by \citet{Marcy51peg}, we decided to 
start a follow-up of some of the stars with planetary companions announced by other teams. 
Confirming and cross-checking the claimed detections is very important, especially when the signal is 
close to the detection limits. 

In this section, we present our radial-velocity measurements for \object{Ups\,And}, \object{55\,Cnc}, \object{47\,UMa}, \object{70\,Vir} and 
\object{{\footnotesize HD}\,187123}.  These stars host one or several planetary companions detected by the 
{\em California \& Carnegie Planet Search} team.  We confirm the orbital solutions for \object{Ups\,And}, \object{70\,Vir} and 
\object{{\footnotesize HD}\,187123}. We are not able to fully confirm the published solutions for \object{55\,Cnc} and \object{47\,UMa} 
because of insufficient data (too short measurement span or poorer precision).

\subsection{\object{Ups\,And} (\object{HD\,9826}, \object{HIP\,7513}, \object{GJ\,61})}\label{Upsand}

The presence of 4.61--d planet in orbit around \object{Ups\,And}  was announced by \citet{Butlerupsand1}. Two additional 
longer-period planets were later uncovered \citep{Butlerupsand} 
by the same team in collaboration with the Advanced Fiber-Optic Echelle spectrometer 
\citep[{\footnotesize AFOE, }][]{Brown1994}  planet search team \citep{KorzennikCSSSS}.

71 {\footnotesize ELODIE} radial-velocity measurements have been gathered between {\footnotesize JD}\,=\,2\,450\,319
(August 1996) and {\footnotesize JD}\,=\,2\,452\,886 (September 2003). The mean velocity uncertainty of these measurements is 
8.3\,m\,s$^{\rm -1}$. We present in Table~\ref{tabupsand} the 3-planet orbital solution derived from our data. 
We note some small differences between our solution and 
the \citet{Butlerupsand} solution as e.g. a slightly lower period value for \object{Ups\,And\,}c and a slightly longer period for 
\object{Ups\,And\,d}.

The {\footnotesize ELODIE} radial velocities and orbital solution are shown in Fig.~\ref{figupsand}. The residuals to the 
fitted orbit are clearly too large as it can be seen from the $\chi^{\rm 2}_{\rm red}$ value: 4.34. 
The residuals obtained with the orbital parameters fixed to the \citet{Butlerupsand} values are even worse: 22.4\,m\,s$^{\rm -1}$ 
($\chi^{\rm 2}_{\rm red}$\,=\,7.75)
We have no explanation for these facts but  
we are nevertheless able to confirm the presence of the three planets around \object{Ups\,And}.

Adopting the same primary mass as in \citet{Butlerupsand}, $M_{\rm 1}$\,=\,1.3\,M$_{\rm sun}$, we compute the planetary 
minimum masses: 
$m_{\rm b}\sin i$\,=\,0.75\,M$_{\rm Jup}$; $m_{\rm c}\sin i$\,=\,2.25\,M$_{\rm Jup}$; 
$m_{\rm d}\sin i$\,=\,3.95\,M$_{\rm Jup}$. The semi-major axes are: $a_{\rm b}$\,=\,0.059\,AU; $a_{\rm c}$\,=\,0.82\,AU; 
 $a_{\rm d}$\,=\,2.57\,AU.

\subsection{\object{55\,Cnc} (\object{HD\,75732}, \object{HIP\,43587}, \object{GJ\,324\,A})}\label{55cnc} 

\begin{figure}[th!]
  \psfig{file=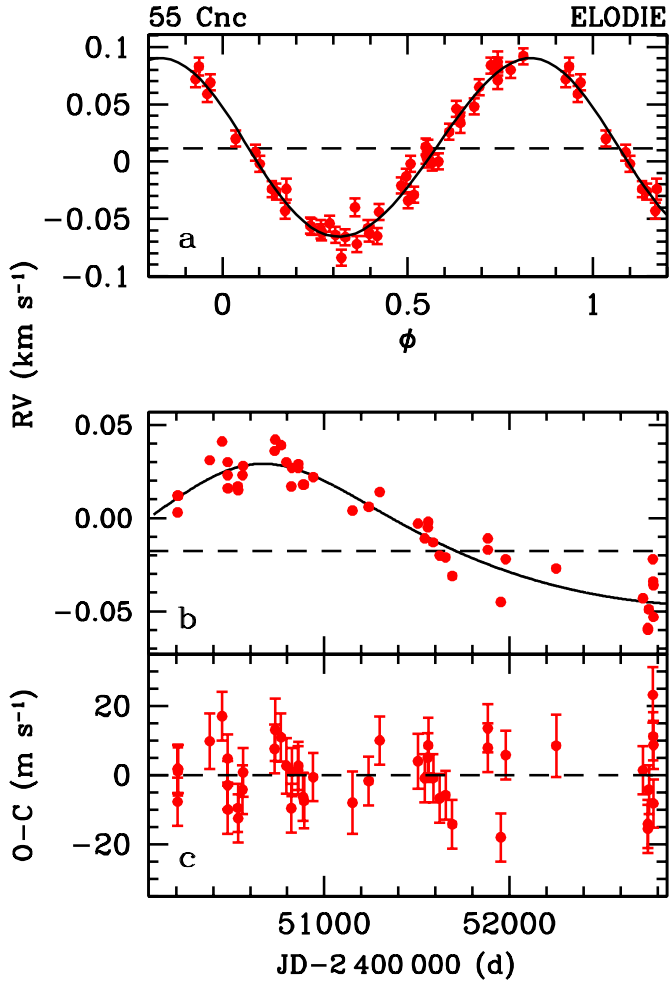,width=\hsize}\\
  \caption{
  \label{fig55cnc}  ELODIE radial-velocity data and orbital solution for \object{55\,Cnc}.
  {\bf a:} Phased-folded velocities for \object{55\,Cnc\,b}. The signal induced by the outer planet is removed.
  {\bf b:} Temporal velocities for \object{55\,Cnc\,d}. The signal of the inner planet is removed.
  {\bf c:} Residuals around the solution.
  }  
\end{figure}

\object{55\,Cnc\,b}, a 14.6--d planet was announced by \citet{Butlerupsand1}. These authors also quoted the presence 
of an additional radial-velocity drift superimposed on the periodic signal. \citet{Marcy55cnc} have recently announced the 
detection of two additional signals with periods of 44.3\,d and 14.7\,yr. The interpretation of the 44.3--d signal 
is uncertain since it corresponds to the rotational period of the star. It thus could be due to stellar intrinsic effects. 
The 14.7--yr signal is interpreted as the result of the presence of a 4\,M$_{\rm Jup}$ planet more than 5\,AU away from the 
star.

48 {\footnotesize ELODIE} radial-velocity measurements of \object{55\,Cnc}  have been obtained between 
{\footnotesize JD}\,=\,2\,450\,207 (May 1996) and {\footnotesize JD}\,=\,2\,452\,779 (May 2003). They have a mean 
uncertainty of 7.3\,m\,s$^{\rm -1}$. Our measurement time span (2571\,d) is shorter than the announced long period by a factor of 2. 
Also, the radial-velocity semi-amplitude of the 44.3--d signal is only 13\,m\,s$^{\rm -1}$.  
These features make difficult the 
characterization of the outer planets with the {\footnotesize ELODIE} measurements but they however allow us to 
 partially confirm 
the orbital solution presented in \citet{Marcy55cnc}. 

We present in Table~\ref{tabupsand} a 2-Keplerian orbital solution for \object{55\,Cnc}. This solution does not 
account for the intermediate signal. We are able to fit a 44--d orbit to the residuals around this solution but 
the fit parameters are too poorly constrained to be significant. Thus, they do not appear in this paper. 
Our data are insufficient for confirming the presence of the intermediate planet.
Figure~\ref{fig55cnc} shows the fitted 
orbits to our velocities as well as the residuals around our solution.

Assuming a primary mass of 0.95\,M$_{\sun}$ for \object{55\,Cnc}, we compute the 
minimum masses of the two companions: $m_{\rm b}\sin i$\,=\,0.91\,M$_{\rm Jup}$ and $m_{\rm d}\sin i$\,=\,2.89\,M$_{\rm Jup}$. 
The orbital semi-major axes are: $a_{\rm b}$\,=\,0.115\,AU and $a_{\rm d}$\,=\,5.28\,AU.

\subsection{\object{47\,UMa} (\object{HD\,95128}, \object{HIP\,53721}, \object{GJ\,407})}\label{47uma}

\begin{figure}
  \psfig{file=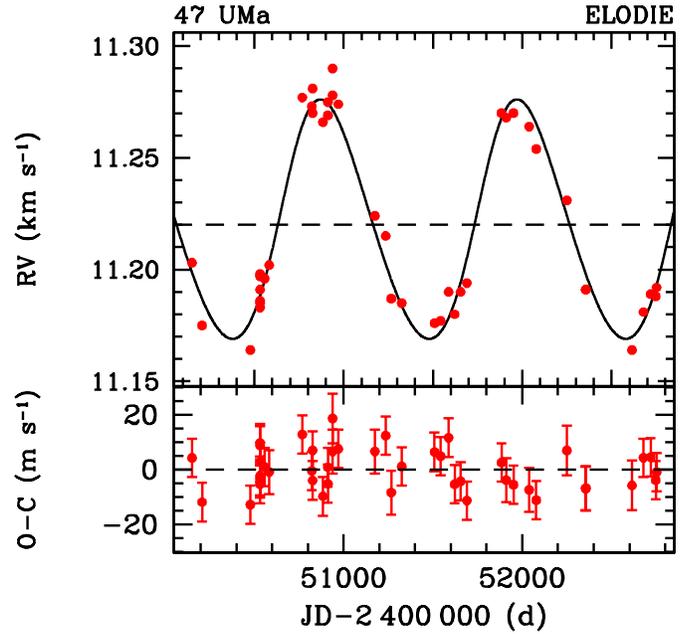,width=\hsize}\\
  \caption{
  \label{fig47uma} {\bf Top:} ELODIE temporal velocities and fitted orbital solution for \object{47\,UMa}.
  {\bf Bottom} Residuals to the  solution.
  }  
\end{figure}

\begin{table*}[t!]
\caption{
\label{tab47uma}
ELODIE orbital solution for \object{47\,UMa}, \object{70\,Vir} and \object{HD\,187123}.
The parameters have the same definitions as in Table~\ref{tab74156}.
}
\begin{tabular}{ll|r@{\,$\pm$\,}lr@{\,$\pm$\,}lr@{\,$\pm$\,}l}
\hline\hline
\noalign{\vspace{0.05cm}}
\multicolumn{2}{c}{}                                                   & \multicolumn{2}{c}{\object{47\,UMa\,b}} &\multicolumn{2}{c}{\object{70\,Vir\,b}} & \multicolumn{2}{c}{\object{HD\,187123\,b}}\\
\hline
\noalign{\vspace{0.05cm}}
$P$                                 & (d)                              & 1100.8  & 7.2               & 116.689    & 0.011          & 3.0966      & 0.0001\\
$T$                                 & (JD$^{\dagger}$)                 & 52\,915 & 64                & 48\,990.39 & 0.33           & 51\,010.972 & 0.027\\
$e$                                 &                                  & 0.097   & 0.039             & 0.397      & 0.005          & \multicolumn{2}{c}{0 (fixed)}\\
$\gamma$                            & (km\,s$^{\rm -1}$)               & 11.220  & 0.001             & 4.951      & 0.001          & $-$16.986$^{\ddagger}$ & 0.002\\
$w$                                 & ($\degr$)                        & 299.7   & 20.3              & 359.40     & 0.92           & \multicolumn{2}{c}{0 (fixed)}\\ 
$K_{\rm 1}$                         & (m\,s$^{\rm -1}$)                & 53.6    & 1.9               & 314.1      & 2.0            & 71.0        & 1.7\\
Slope                               & (m\,s$^{\rm -1}$\,yr$^{\rm -1}$) & \multicolumn{2}{c}{--}      & \multicolumn{2}{c}{--}      & $-$8.8      & 0.9\\
$a_{\rm 1} \sin i$                  & (10$^{\rm -5}$AU)                & 540     & 20                & 309        & 2              & 2.01        & 0.06\\
$f_{\rm 1}(m)$                      & (10$^{\rm -10}$M$_{\sun}$)       & 173     & 19                & 2897       & 55             & 1.14        & 0.10\\
$m\sin i$                           & (M$_{\rm Jup}$)                  & 2.76    & 0.10              & 6.56       & 0.04           & 0.52        & 0.01\\
$a$                                 & (AU)                             & \multicolumn{2}{c}{2.11}    & \multicolumn{2}{c}{0.456}   & \multicolumn{2}{c}{0.042}\\[0.05cm]
\hline
\noalign{\vspace{0.05cm}}
$N$                                 &                                  & \multicolumn{2}{c}{44}      & \multicolumn{2}{c}{35}      & \multicolumn{2}{c}{57}\\ 
$\sigma_{\rm O-C}$                  & (m\,s$^{\rm -1}$)                & \multicolumn{2}{c}{7.4}     & \multicolumn{2}{c}{6.1}     & \multicolumn{2}{c}{10.5}\\
$\chi^{\rm 2}_{\rm red}$            &                                  & \multicolumn{2}{c}{1.19}    & \multicolumn{2}{c}{0.87}    & \multicolumn{2}{c}{1.37}\\[0.05cm]
\hline
\end{tabular}

$^{\dagger}$=\,{\footnotesize JD}$-$2\,400\,000, $^\ddagger$ at {\footnotesize JD}\,=\,2\,451\,500
\end{table*}

The detection of a long-period planet in orbit around \object{47\,UMa}  was announced  in \citet{Butler47uma}. 
A preliminary {\footnotesize ELODIE} orbit for this companion was published in \citet{Naefcssss}.
More recently, \citet{Fischer47umac} have presented an updated 
orbital solution including another planetary companion on a longer-period orbit.

We have obtained 44 {\footnotesize ELODIE} radial-velocity measurements of \object{47\,UMa} between 
{\footnotesize JD}\,=\,2\,450\,150 (March 1996) and {\footnotesize JD}\,=\,2\,452\,752 (April 2003). The mean 
velocity uncertainty of these measurements is 7.3\,m\,s$^{\rm -1}$. We are unable with our data to confirm the 
presence of the second planetary companion. The solution presented in Table~\ref{tab47uma} only 
includes one Keplerian orbit. The residuals to this solution (7.4\,m\,$s^{\rm -1}$) are in good agreement with our 
uncertainties. We do not see any clear velocity trend in these residuals (Fig.~\ref{fig47uma}). With the same 
primary mass as in \citet{Fischer47umac}, $M_{\rm 1}$\,=\,1.03\,M$_{\sun}$, we compute the planet minimum mass and its 
orbital semi-major axis: $m_{\rm 2}\sin i$\,=\,2.76\,M$_{\rm Jup}$ and $a$\,=\,2.11\,AU.

We have tried without success to adjust a 2-Keplerian orbital solution to our velocities, starting from the solution 
published in \citet{Fischer47umac}. Moreover, the residuals we obtain by fixing the \citet{Fischer47umac} solution 
to our data are large: $\sigma_{\rm O-C}$\,=\,13\,m\,s$^{\rm -1}$. This solution is thus not compatible with the 
{\footnotesize ELODIE} data. In a recent paper, \citet{Ford2003} has shown that the orbital solution for 
\object{47\,UMa} was almost unconstrained by the \citet{Fischer47umac} velocities. For example, he has obtained 
a statistically equivalent solution with a period three times longer and an eccentricity eight times larger than the 
published values. At this point, the {\footnotesize ELODIE} data are not sufficient for confirming or ruling out the 
presence of another body in the system. They are well fitted by a single planet orbit. These facts demonstrate that the 
\citet{Fischer47umac} orbit is very preliminary.

\subsection{\object{70\,Vir} (\object{HD\,117176}, \object{HIP\,65721}, \object{GJ\,512.1})}\label{70vir}

The detection of a massive planet on a moderately  eccentric 116--d orbit around \object{70\,Vir}  was announced in \citet{Marcy70Vir}. A preliminary 
{\footnotesize ELODIE} orbit confirming the \citet{Marcy70Vir} solution was presented in \citet{Naefcssss}.

\begin{figure}
  \psfig{file=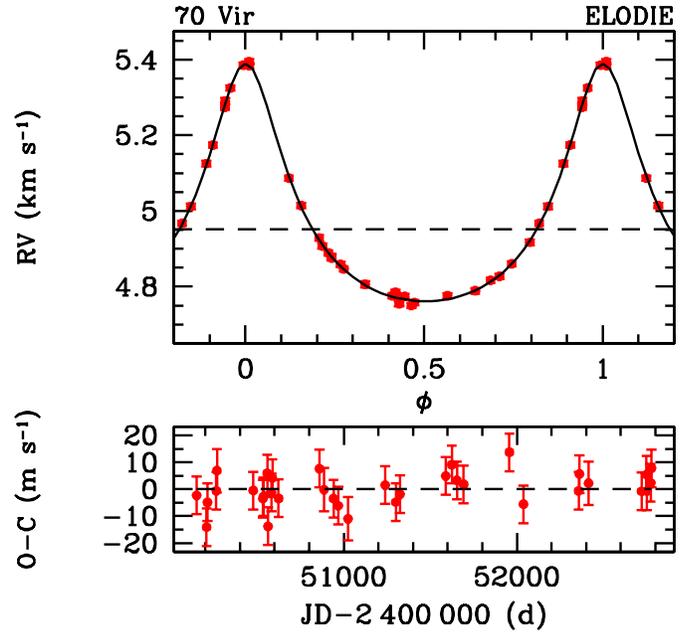,width=\hsize}\\
  \caption{
  \label{fig70vir} ELODIE radial-velocity data and orbital solutions for \object{70\,Vir}. 
  {\bf Top:} Phased-folded velocities.
  {\bf Bottom:} Residuals to the fitted orbital solution.
  }  
\end{figure}

We now have 35 {\footnotesize ELODIE} radial-velocity measurements. They have been obtained between 
{\footnotesize JD}\,=\,2\,450\,150 (March 1996) and {\footnotesize JD}\,2\,452\,777 (May 2003). They have a mean velocity 
uncertainty of 7.1\,m\,s$^{\rm -1}$. We present in Table~\ref{tab47uma} the orbital solution fitted to our velocities.
This solution is in full agreement with the \citet{Marcy70Vir} solution. It is also much more precise than our preliminary 
solution \citep{Naefcssss}. The improvement is due to the use of the new cross-correlation template reducing the influence of 
the atmospheric telluric lines \citep{Pepe2002}. With residuals of 6.1\,m\,s$^{\rm -1}$, it is to date the most precise 
published {\footnotesize ELODIE} orbital solution and it thus illustrates well the level of precision presently achieved with 
this instrument. Figure~\ref{fig70vir} shows our radial velocities and the fitted solution.

With the primary mass quoted in \citet{Marcy70Vir}, $M_{\rm 1}$\,=\,0.92\,M$_{\sun}$, we find a minimum mass of 
6.56\,M$_{\rm Jup}$ for 70\,Vir\,b. Its orbital semi-major axis is 0.456\,AU.

\subsection{\object{HD\,187123} (\object{HIP\,97336})}\label{187123}

The presence of a 3.1--d planet in orbit around \object{{\footnotesize HD}\,187123}  was announced by 
\citet{Butler187123}. An updated orbital solution including an additional linear velocity drift was presented in 
\citet{Vogt6planets}.

\begin{figure}
  \psfig{file=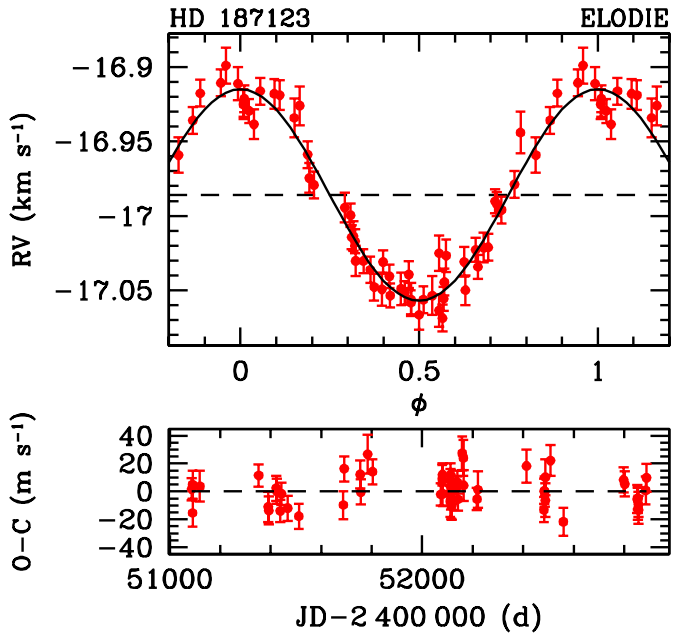,width=\hsize}\\
  \caption{
  \label{fig187123} ELODIE radial-velocity data and orbital solutions for \object{HD\,187123}. 
  {\bf Top:} Phased-folded velocities. The velocities are corrected for a linear drift of $-$8.8\,m\,s$^{\rm -1}$\,yr$^{\rm -1}$.
  {\bf Bottom:} Residuals to the fitted orbit.
  }  
\end{figure}

We started our {\footnotesize ELODIE} measurements of \object{{\footnotesize HD}\,187123} shortly after the discovery announcement.
We now have 57 radial-velocity measurements of this star spanning from {\footnotesize JD}\,2\,451\,087 (September 1998) to 
{\footnotesize JD}\,2\,452\,889 (September 2003). They have a mean velocity uncertainty of 9.6\,m\,s$^{\rm -1}$. The orbital 
solution fitted to this velocities is presented in Table~\ref{tab47uma}. We fitted a circular orbit since our data do not 
exhibit a significant eccentricity according to the \citet{Lucy} test. Our solution also includes an additional linear 
velocity drift. The derived parameters are in full agreement with those 
presented in \citet{Vogt6planets}. We therefore confirm the presence of the short-period planet and the linear drift 
probably due to another massive body in the system. Figure~\ref{fig187123} shows the {\footnotesize ELODIE} velocities and 
the fitted orbital solution.

Using the primary published in \citet{Santosstat}, $M_{\rm 1}$\,=\,1.05\,M$_{\sun}$, we derive 
$m_{\rm 2}\sin i$\,=\,0.52\,M$_{\rm Jup}$ and $a$\,=\,0.042\,AU for \object{{\footnotesize HD}\,187123\,b}.

\section{Summary}\label{conc}

In this paper, we have presented our {\footnotesize ELODIE} radial-velocity data and derived planetary solutions for
\object{{\footnotesize HD}\,74156}, \object{14\,Her}, \object{{\footnotesize HD}\,209458}, \object{51\,Peg}, 
\object{Ups\,And}, \object{55\,Cnc}, \object{47\,UMa}, \object{70\,Vir} and 
\object{{\footnotesize HD}\,187123}, including three new {\footnotesize ELODIE} planets, updated orbital solutions of 
previously published candidates and confirmations of detections from other planet search programmes.

The radial-velocity variations detected for \object{{\footnotesize HD}\,74156} are found to be due to the 
presence of a planetary system consisting of a 1.86\,M$_{\rm Jup}$ planet at 0.294\,AU and a 6.17\,M$_{\rm Jup}$ planet 
on a 5.5--yr orbit.
For \object{14\,Her}, a 4.74\,M$_{\rm Jup}$ planet at 2.80\,AU induces the main detected signal. An additional
linear velocity drift revealing the presence of another not yet characterized massive body in the 
system is detected as well for \object{14\,Her}.

We have	presented updated more precise orbits for \object{{\footnotesize HD}\,209458} and \object{51\,Peg}, two stars hosting short-period 
planets detected with {\footnotesize ELODIE}.

We have confirmed the previously published orbital solutions for \object{Ups\,And}, \object{70\,Vir} and \object{{\footnotesize HD}\,187123}.
We have also partially confirmed the solutions for \object{55\,Cnc} and \object{47\,UMa}. The velocity signals of the inner and the outer planet 
orbiting \object{55\,Cnc} are clearly detected. 
We have no evidence for the presence of \object{47\,UMa\,c}. Our data is not compatible with the \citet{Fischer47umac} solution.
However, the radial-velocity semi-amplitude induced by this object (11\,m\,s$^{\rm -1}$) is small 
compared to our measurement precision (7.3\,m\,s$^{\rm -1}$). We thus cannot rule-out this claimed detection.

\begin{acknowledgements}
We acknowledge support from the Swiss National Research Found 
({\footnotesize FNRS}), the Geneva University and the French 
{\footnotesize CNRS}. We are grateful to the Observatoire de 
Haute-Provence for the generous time allocation. 
This research has made use 
of the {\footnotesize SIMBAD} database, operated at 
{\footnotesize CDS}, Strasbourg, France.
\end{acknowledgements}

\bibliographystyle{aa}
\bibliography{naef_eloIII}
%\listofobjects
\end{document}